\documentclass[10pt,fleqn]{article}
\usepackage[german,british]{babel}
\usepackage{cite}

\usepackage{fancyheadings}

\usepackage{amsmath}
\usepackage{amssymb}

\newcommand{\dif}{\mathrm{d}}

\begin{document}

\title{On astrophysical bounds of the cosmological constant}

\author{
Andr\'{e}s Balaguera-Antol\'{\i}nez\footnote{e-mail: {\tt a-balagu@uniandes.edu.co}}~,
Christian G. B\"ohmer\footnote{e-mail: {\tt boehmer@hep.itp.tuwien.ac.at}} \and 
and Marek Nowakowski\footnote{e-mail: {\tt mnowakos@uniandes.edu.co}} \\[2ex]
\footnotemark[1]~\footnotemark[3]
Departamento de F\'{\i}sica, Universidad de los Andes,\\ 
A.A. 4976, Bogot\'a, D.C., Colombia. \\[2ex]
\footnotemark[2]
Institut f\"ur Theoretische Physik, Technische Universit\"at Wien,\\ 
Wiedner Hauptstr. 8-10, A-1040 Wien, Austria.
}

\date{}

\maketitle

\thispagestyle{fancy}
\setlength{\headrulewidth}{0pt}
\rhead{TUW--04--23}

\begin{abstract}
Astrophysical bounds on the cosmological constant are examined
for spherically symmetric bodies. Similar limits emerge from
hydrostatical and gravitational equilibrium and the validity of
the Newtonian limit. 
The methods in use seem to be disjoint from the basic principles, 
however they have the same implication regarding the upper bounds.
Therefore we will compare different inequalities and comment on 
the possible relationship between them. 
These inequalities are of relevance for the so-called coincidence problem
and for the bound of the cosmological constant which comes surprisingly close to the
`experimental' value.
\end{abstract}

\mbox{} \\ 
\mbox{} 
\mbox{} \\
\mbox{} 

\newpage
\section{Introduction}

The cosmological constant $\Lambda$ has a history as long as 
General Relativity itself: introduced to make the universe 
static, abandoned after the discovery of the Hubble expansion 
and introduced recently to explain the acceleration of the 
Universe~\cite{Krauss:1998,Perlmutter:1999jt}. It is true 
that the latter phenomenon can, in principle,  be also accounted 
for by other explanations like quintessence~\cite{Ratra:1987rm,Wetterich:1987fm}, 
Chaplygin Gas~\cite{Kamenshchik:2001cp}, mixture 
models~\cite{Kremer:2002hz,Kremer:2003ev,Kremer:2003vs,Kremer:2004tw} 
and modified gravity~\cite{Nojiri:2003} etc, but the simplest 
explanation for an acceleration can be obtained from a positive 
cosmological constant with
\begin{align}
      \Lambda &= 8\pi G\rho_{{\rm vac}} \,,
      \label{cc1} \\
      \rho_{{\rm vac}}&\approx (0.7-0.8)\rho_{{\rm crit}} \,.
      \nonumber
\end{align}
This value is surprisingly close to a prediction made long ago by 
Zel'dovich~\cite{Zeldovich:1967gd,Carneiro:2002iv,Tegmark:2000qy,Matthews:1988}, namely,
\begin{align}
      \label{cc2}
      \Lambda\approx G^{2} m^{6}\,,
\end{align}
where $m$ is a typical hadronic mass scale ($0.15-1$ GeV). Indeed, 
combining this with $H_{0}=m^{3}$
\cite{Eddington:1931,MenaMarugan:2001qn}, 
one gets the result that the Hubble parameter is dominated by 
the cosmological constant $\Lambda$.

Cosmological observations~\cite{Krauss:2003yb,Perlmutter:1998zf,Riess:2004nr} 
give strong indications of the presence  of a positive cosmological constant 
which would mean that the universe is of \mbox{de Sitter} type.
Moreover, recent studies~\cite{Allen:2004cd} of X-Rays hint toward a constant 
density of the \emph{dark energy} which is thought responsible 
for the acceleration, also in agreement with a positive $\Lambda$. 
The use of the different distance measurements in a $\Lambda$CDM 
model (models of cold dark matter with $\Lambda$) 
also shows strong evidence for a positive cosmological constant
($0.47<\Omega_{\Lambda}<0.82,\Omega_{\Lambda}=\rho_{{\rm vac}}/\rho_{{\rm crit}}$)~\cite{Kunz:2004ry}.

Such a relatively large cosmological constant has astrophysical
implications. This was realized as early as 1939 by Gamow and 
Teller~\cite{Gamow:1939}. The situation is best paraphrased 
in~\cite{North:1990} which summarizes the work of several 
authors in the $30$'s and $40$'s: ``The essential difficulty with a relativistic theory in 
which $\lambda$ [the Cosmological Constant] is positive is 
that of accounting for the formation and condensation in terms 
of gravitational instability; for, to use the \emph{`force'} 
metaphor, the present expansion indicates that the force of 
cosmic repulsion exceeds those of gravitational 
attraction~\cite{Gamow:1939}. This is not likely to disturb 
the stability of systems (such as the galaxy) of high average 
density, but it is likely to prevent new condensation in regions of low
density.''

Indeed, several studies have corroborated this 
conclusion and many well known systems are 
reconsidered with the inclusion of the formerly omitted $\Lambda$-term. 
Another, more mathematical reason to include $\Lambda$ is the richer asymptotic 
structure of these spacetimes and moreover the AdS/CFT 
correspondence~\cite{Maldacena:1998re} motivating much research.

Examples of recent research are the Einstein-Yang-Mills-$\Lambda$ system~\cite{Volkov:1996qj},
$(n+4)$ dimensional EYM-$\Lambda$ theory~\cite{Brihaye:2004qr},
perfect fluid solutions with cosmological 
constant~\cite{Boehmer:2002gg,Boehmer:2003uz,Stuchlik:2000,Weyl:1919},
consistency with the Newtonian limit~\cite{Nowakowski:2000dr,Sussman:2003km},
gravitational equilibrium via the virial 
theorem~\cite{Nowakowski:2001zw} and also Einstein-Cartan and Einstein-Cartan-Dirac
theory with cosmological constant~\cite{Boehmer:2003iv,Dimakis:1985jb}. 
It is remarkable that these examples have something in common, 
namely, in each of them there exists an upper bound of the cosmological
constant.

Gravitational equilibrium
and perfect fluid considerations surprisingly predict the same
upper bound in terms of the mean density of an astrophysical object,
$\Lambda < 4\pi G \bar{\rho}$. Hence it seems that there
is some relation between the existence of exact solutions of Einstein's
field equations and the virial theorem. In the EYM-$\Lambda$ system,
one finds $\Lambda < \Lambda_{{\rm crit}}$.
In the $(n+4)$ dimensional EYM-$\Lambda$ theory 
it was found that the cosmological constant is bounded, 
such that  $\Lambda_{(n+4)} < g^2/ 2 G_{(n+4)}$ and constant
density solutions exist for $\Lambda < \Lambda_{{\rm S}}$,
for larger values they are unphysical. 
Finally we mention that supergravity~\cite{VanNieuwenhuizen:1981ae}
even forbids a positive value and therefore predicts $\Lambda\leq 0$.
It should be noted that a lower bound never occurs.

In~\cite{Nowakowski:2001zw} it was found that the more 
the shape of the object deviates from spherical
symmetry, the more difficult it is to reach the 
gravitational equilibrium in presence of a positive 
cosmological constant  
for low density astrophysical objects. This result 
was based on the assumption of gravitational equilibrium. Of 
course, equally valid and useful concept is the hydrodynamical 
equilibrium combined with Einstein's equations. These two concepts 
of equilibrium give us information on $\Lambda$ in terms of 
inequalities. A third source of useful information on $\Lambda$ is the 
Newtonian limit, which also gives us an inequality.

In this paper we will compare different inequalities and point 
out some interesting coincidences. We will use these results 
to infer on certain static properties of astrophysical objects.
Unless otherwise stated we put $G=c=\hbar=1$.

\section{Newtonian hydrostatic equilibrium}

In order to be self consistent, some basic relations are reviewed
as shortly as possible. 

In Newtonian astrophysics a spherically symmetric  object is stable if the 
gravitational and matter pressure are in equilibrium.
This condition leads to
\begin{align}
      P'(r) = -r\rho(r)\Bigl(\frac{m(r)}{r^3}
      -\frac{\Lambda}{3}\Bigr)\,,
      \label{eq:newton1}
\end{align}   
which is sometimes called~\cite[p.301]{Weinberg:1972} the ``fundamental equation of Newtonian
astrophysics''. The mass function is as usual defined by
\begin{align}
      m(r) = \int_0^r 4\pi \rho(s) s^2 \dif s \,.
      \label{eq:mass}
\end{align}  
Furthermore let the mean density up to $r$ be defined by
\begin{align}
      \bar{\rho}(r) = \frac{3}{4\pi}\frac{m(r)}{r^3}\,,
      \label{eq:rhobar}
\end{align}  
which then has the following properties
\begin{align}
      \bar{\rho}_c = \rho_c\,, \qquad \bar{\rho}(r) \geq \rho(r)\quad\forall\ r \,,
      \label{eq:rhobar1}
\end{align}
where $\rho_c$ is the central density.
If a density profile $\rho(r)$ for the astrophysical object is prescribed, one
can firstly use~(\ref{eq:mass}) to find the mass function and secondly
integrate~(\ref{eq:newton1}) to find the pressure $P(r)$. 

However the most physical starting point is to prescribe an equation of state $\rho=\rho(P)$. 
First we rewrite~(\ref{eq:newton1}) with the help of~(\ref{eq:rhobar})
and find
\begin{align}
      P'(r) = -r\frac{\rho(r)}{3}
      \Bigl(4\pi\bar{\rho}(r)-\Lambda\Bigr)\,.
      \label{eq:newton2}
\end{align}
Then, differentiating the mean density~(\ref{eq:rhobar})
leads to
\begin{align}
      \bar{\rho}'(r)=\frac{3}{r}\bigl(\rho(P(r))-\bar{\rho}(r)\bigr)\,.
      \label{eq:newton3}
\end{align}
Thus the two equations~(\ref{eq:newton2}) and~(\ref{eq:newton3})
form a system of differential equations in $P(r)$ and $\bar{\rho}(r)$. 

For any physically reasonable astrophysical object, the pressure 
and density must be monotonically decreasing functions 
of the object's radius. Hence negativity of $P'(r)$ 
from~(\ref{eq:newton2}) implies
\begin{align}
      \Lambda < 4\pi\bar{\rho}_b \,,
      \label{eq:newton4}
\end{align} 
since by virtue of~(\ref{eq:rhobar1}) the mean density is a
monotonically decreasing function having its minimum at the
boundary of the stellar object, where it is just the mean
density of the stellar object. The boundary is defined by the
radius $R$ for which the pressure vanishes, $P(R)=0$.

If, for the moment. we assume $\rho(r)=\rho =\mbox{const.}$,  then~(\ref{eq:newton2})
gives
\begin{align}
      P'(r) = -r\frac{\rho}{3}
      \Bigl(4\pi\rho-\Lambda\Bigr)\,,
      \label{eq:newton5}
\end{align}
and is, as before, well defined if and only if $\Lambda <4\pi\rho$.
However, if $\Lambda=4\pi\rho$ we find that the pressure must be
constant for all $r$. Hence one may say that this is the Newtonian
analogue of the Einstein static universe. Note that the cosmological
constant in the Newtonian case is independent of the pressure as 
is expected from general grounds since in Newton's theory of gravity
pressure does not contribute to the energy momentum tensor.

Moreover if the pressure increases near the center, it automatically
diverges. This changes considerably in the general relativistic case.

\section{General relativistic hydrostatic equilibrium}

Most parts of the former section can still be applied to the
general relativistic case. The only difference is that the
differential equation for the pressure~(\ref{eq:newton2})
is replaced by the Tolman-Oppenheimer-Volkoff 
equation~\cite{Oppenheimer:1939ne,Tolman:1939jz}
with cosmological constant (TOV-$\Lambda$) and is given as
\begin{align}
      P'(r) =-r\frac{\rho(r)}{3}\Bigl(1+\frac{\lambda P(r)}{\rho(r)}\Bigr)
      \left(\frac{\lambda 12\pi P(r) + 4\pi\bar{\rho}(r)-\Lambda} 
      {1-\lambda\frac{8\pi}{3}\bar{\rho}(r)r^{2}-\lambda\frac{\Lambda}{3}r^{2} }\right)\,,
      \label{eq:tov}
\end{align}
where the velocity of light has been included by $\lambda = 1/c^2$ 
to facilitate the comparison with the Newtonian case.
The Newtonian limit~\cite{Ehlers:1991}\nocite{Ferrarese:1991},  
$\lambda\rightarrow 0$, of~(\ref{eq:tov}) reproduces~(\ref{eq:newton2}). 
We will use geometrical units henceforth.

The TOV-$\Lambda$ equation together with~(\ref{eq:newton3}) 
form a system of differential equations if an equation of
state $\rho=\rho(P)$ is specified. 

The right hand side of~(\ref{eq:tov})
is well defined if the denominator is
\begin{align}
      y^2 := 1-\frac{8\pi}{3}\bar{\rho}(r)r^{2}-\frac{\Lambda}{3}r^{2}
      \geq 0 \qquad\forall\ r\,,
      \label{eq:gen1}
\end{align} 
where we introduced a new variable $y$.
The interesting question is whether one can find a general relativistic
analogue of the upper bound of the cosmological constant~(\ref{eq:newton4}).
The answer to this question is affirmative, and it is related to fact that 
the TOV-$\Lambda$ equation be well defined.

One can derive~\cite{Buchdahl:1959,Rendall:1991hg,Boehmer:2002gg} the following inequality 
\begin{align}
      y \geq \frac{12\pi P(r)+ 4\pi \bar{\rho}(r)-\Lambda}
      {12\pi P_{c}+ 4\pi\bar{\rho}_{c}-\Lambda}\,,
      \label{eq:gen2}
\end{align}
which can be used to arrive at the desired equation. Since the
boundary of any astrophysical object is defined by $P=0$, 
the variable $y$ is well defined up to the boundary if
the remaining terms of the numerator satisfy
\begin{align}
      \Lambda < 4\pi \bar{\rho}_b \,.
      \label{eq:gen3}
\end{align}
Therefore, the Newtonian upper bound of the cosmological 
constant could exactly be reproduced in the general
relativistic case. 

Before continuing, let us shortly review qualitatively how 
one derives~(\ref{eq:gen2}). The Einstein field equations
for a static and spherically symmetric perfect fluid are
three independent differential equations for four free
functions. Hence one can specify one of them. The most
physical one is to prescribe an equation of state $\rho=\rho(P)$.
The TOV-$\Lambda$ equation can be extracted from this system
by eliminating the free function $g_{rr}$ if the line element
is given in its usual form. 
\footnote{$\dif s^{2}=-e^{\nu (r)}\dif t^{2}+e^{a(r)}
\dif r^2+r^{2}(\dif\theta^{2}+\sin^{2}\negmedspace\theta\,\dif\phi^{2})$.}
On the other hand, one can also eliminate the pressure from
the second field equation and the conservation equation.
After introducing Buchdahl variables~\cite{Buchdahl:1959}
and assuming that the pressure is a decreasing function near the center
and prescribing a monotonic equation of state one arrives at~(\ref{eq:gen2}).
Without the cosmological constant one can directly read 
it off from~\cite{Buchdahl:1959,Rendall:1991hg}. The cosmological constant is
incorporated by using the standard substitution 
$\rho\rightarrow\rho + \Lambda/8\pi$ and $P\rightarrow P - \Lambda/8\pi$ 
which leads again to equation~(\ref{eq:gen2}). For an explicit derivation with
the cosmological constant see~\cite[p.42--p.47]{Boehmer:2002gg}.

One must be slightly careful with the above, since in case of
a constant density distribution $\bar{\rho}=\rho_0$ one can integrate the 
TOV-$\Lambda$ equation exactly and no
upper bound $\Lambda=4\pi\bar{\rho}$ arises, which is
expected since the Einstein static universe with non-vanishing
pressure is also a solution of the field equations. The interior 
metric has the geometry of a three sphere which is
conformally flat~\cite{Stephani:1967} and one just
has to deal with a harmless coordinate singularity.
See~\cite{Boehmer:2003uz} for a complete discussion of 
constant density solutions with cosmological constant where it
is also shown that in contrast to the Newtonian case one can
construct solutions with increasing pressure near the center.

\section{Buchdahl inequalities}
\label{se:buch}

There are essentially two ways of solving the field equations, as already
mentioned in the previous section. With the help of Buchdahl 
variables~\cite{Buchdahl:1959} one has a very convenient scheme to 
compare a general monotonically decreasing solution with a constant 
density solution. The latter is defined by the boundary of the mean 
density of the general solution.

To perform the described procedure, Buchdahl~\cite{Buchdahl:1959} 
originally assumed the existence of a global solution, which was
later proven to exist in~\cite{Rendall:1991hg}. Up to 
$\Lambda<4\pi \bar{\rho}_b$ one can use the methods of~\cite{Rendall:1991hg}
to prove the existence of a global solution with non-vanishing
cosmological constant~\cite[Theorem~5]{Boehmer:2002gg}. 

The derivation of the Buchdahl inequality is quite involved,
therefore we only state the result and refer 
to~\cite{Buchdahl:1959} and~\cite[p.367]{Beig:2000yf} without 
and~\cite[Theorem~6]{Boehmer:2002gg} with the cosmological constant.
The generalized Buchdahl inequality is given by
\begin{align}
      \sqrt{1-\frac{8\pi}{3}\bar{\rho}_b R^{2}
      -\frac{\Lambda}{3}R^{2}}  
      \geq \frac{1}{3}-\frac{\Lambda}{12\pi\bar{\rho}_b}\,.
      \label{eq:buch1}
\end{align}
For constant density solutions, generalized Buchdahl inequalities
can be found in~\cite{Mak:2001gg,Boehmer:2003uz}. In~\cite{Mak:2001gg}
the surface redshift with $\Lambda$ was derived from~(\ref{eq:buch1}).

For $\Lambda=0$ this inequality reduces to
\begin{align}
      \sqrt{1-\frac{8\pi}{3}\bar{\rho}_b R^{2}}\geq \frac{1}{3}\,,
      \label{eq:buch2}
\end{align}
which with $M=(4\pi/3)\bar{\rho}R^3$ leads to its most popular form 
\begin{align}
      \sqrt{1-\frac{2M}{R}}\geq \frac{1}{3}\,,\qquad
      2M < \frac{8}{9} R\,, 
      \label{eq:buch3}
\end{align}
from which one reads off the well known result that radii of
static perfect fluid spheres are larger than the Schwarzschild radius
of the corresponding mass.

Let us for the moment assume that a global solution exists independent 
of the cosmological constant, so we assume existence for 
$\Lambda\geq 4\pi\bar{\rho}_b$. From the generalized Buchdahl 
inequality~(\ref{eq:buch1}) one finds instead
\begin{align}
      R^2 \leq 
      \frac{\frac{1}{3}\Bigl(4-\frac{\Lambda}{4\pi\bar{\rho}_{b}}\Bigr)}
      {4\pi\bar{\rho}_{b}}\,,
      \label{eq:buch4}
\end{align}
which after some algebra can be written as
\begin{align}
      \frac{\Lambda}{16\pi\bar{\rho}_{b}} \leq
      1-\frac{9M}{4R}\,.
      \label{eq:buch5}
\end{align}
Since the right-hand side is bounded by $1$, one simply finds
\begin{align}
      \Lambda \leq 16\pi\bar{\rho}_{b}\,,
      \label{eq:buch6}
\end{align}
which is in full agreement with the results from constant density
solutions~\cite{Boehmer:2003uz}, where solutions exist up to a
cosmological constant $\Lambda_{{\rm S}}$, given by
\begin{align}
      \Lambda_{{\rm S}} = 16\pi\bar{\rho}+24 P_c\,.
      \label{eq:buch7}
\end{align}
Therefore it seems that properties of the constant density solutions
in~\cite{Boehmer:2003uz} are also a feature of general solutions with
any given equation of state. However, existence of solutions with
$\Lambda \geq 4\pi\bar{\rho}_{b}$ so far could not be proved. Nonetheless
{\it we conjecture their existence based on the generalized Buchdahl 
inequality}, similar to Buchdahl's original assumption. 
Moreover one may use $\bar{\rho}_b = 3M/(4\pi R^3)$ 
in~(\ref{eq:buch4}) which yields 
\begin{align}
      3M \leq \frac{2}{3}R + 
      R\sqrt{\frac{4}{9}-\frac{\Lambda}{3}R^2}\,,
      \label{eq:buch8}
\end{align}
and reduces to~(\ref{eq:buch3}) for $\Lambda=0$. The above equation
may also be regarded as the generalized Buchdahl inequality since
it gives an upper bound on the mass.
It is tempting to make a Taylor expansion with respect to $\Lambda$ in 
equation~(\ref{eq:buch8}), trying to extract some further
information. However, by looking at~(\ref{eq:buch4}) and~(\ref{eq:buch5})
one sees that nothing can be found  since both equations are
linear in $\Lambda$. Moreover it should be noted that the square root
appearing on the right hand side of~(\ref{eq:buch8}) is well defined if
\begin{align}
       \label{radius}      
       R \leq \sqrt{\frac{4}{3}}\frac{1}{\sqrt{\Lambda}}\,,      
\end{align}
is satisfied by the object's radius $R$. Inserting the 
highest possible radius~(\ref{radius}) in~(\ref{eq:buch8}) we obtain
\begin{align} 
      \label{mass}
      \frac{2}{3}\sqrt{\frac{4}{3}}\frac{1}{\sqrt{\Lambda}} \geq 3M\,.
\end{align}
It is now worthwhile to compare the inequalities~(\ref{radius}) and
(\ref{mass}) with constraints arising from the Newtonian limit. The Newtonian
approximation has, of course, a certain range of validity. In the static,
non-relativistic limit, the requirement of weak fields results into two
strong inequalities of the form
\begin{align} 
      \sqrt{\frac{6}{\Lambda}} \gg R \gg M\,, 
      \qquad
      M_{{\rm max}} = \frac{2}{3}\sqrt{2}\frac{1}{\sqrt{\Lambda}} \gg M\,,
      \label{NL}
\end{align}
valid up to small corrections of the order $M/M_{{\rm max}}$ 
and $(M/M_{{\rm max}})^2$. 
What is striking, is the similarity of inequalities~(\ref{radius}) and
(\ref{mass}) to the restrictions~(\ref{NL}) of the Newtonian limit.
Disregarding different numerical factors which are of the order of unity and the
fact that~(\ref{NL}) are strong inequalities, we see that both are essentially
the same. Inequalities~(\ref{radius}) and~(\ref{mass}) can be understood as
constraints on  $\Lambda$ to keep the object in hydrostatic equilibrium from which
they are derived. On the other hand,~(\ref{NL}) ensures that the gravitational
fields are not too strong. Hence in both cases one expects a restriction on
a positive cosmological constant. What is surprising, however, is the fact that
these restrictions are so similar in both cases. We will discuss further
relevance of equations~(\ref{radius})-(\ref{NL}) in section 8. 

\newcommand{\lp}{\left(}
\newcommand{\rp}{\right)}

\section{Lessons from constant density solutions}
In this section, static and spherically symmetric constant
density perfect fluids are shortly revisited. A new inequality
is derived which in the constant density case is shown to reduce 
to an equality. In the subsequent section this is used to
extract information on the cosmological constant which 
is encoded in a cubic equation in $\Lambda$. 

\subsection{With equation of state}
Assume than a constant density and an equation of state are prescribed, 
for instance, the equation of state of a polytrope 
\begin{align}
      \label{eos1}
      P= \kappa \rho^{\gamma} \,.
\end{align}
In this case we  have $P'(r)=0$, that is, we also have a constant pressure 
(we can take this as an approximation in the sense that  $P'(r)\approx 0$ 
for a slowly varying density or pressure profile). 
Using this in the TOV-$\Lambda$ equation~(\ref{eq:tov}) leads to
\begin{align}
      \label{eos3}
      P'(r) =-\frac{1}{3}r^{2}\lp\rho+P\rp
      \lp \frac{4\pi\bar{\rho}+12\pi P-\Lambda}
      {1-\frac{8\pi}{3}\bar{\rho}r^{2}-\frac{\Lambda}{3}r^{2}}\rp=0\,,
\end{align}
and implies an equation for the constant density $\rho$:
\begin{align}
      \label{eos4}
      4\pi\rho+12\pi\kappa\rho^{\gamma}-\Lambda=0\,.
\end{align}
Note that with vanishing cosmological constant $\Lambda=0$, there would not be 
any acceptable solution for a positive energy density. Thus, we can say that in 
the presence of a cosmological constant, a system with constant density and constant 
pressure can be in hydrostatical equilibrium; the Einstein static universe
is exactly such a system. In the case with $\Lambda=0$, such 
a system could not be in hydrostatical equilibrium; sooner or later the system 
will collapse under it's own gravity. From~(\ref{eos4}) we then can find 
$\rho=\rho(\kappa, \gamma, \Lambda)$.
To illustrate the consequence of equation~(\ref{eos4}) let us assume that $\gamma=1$.
It then follows that an astrophysical object with $\rho \sim \rho_{\rm crit}$ is stable.
Such stable objects would have the lowest possible density among stable astrophysical objects and could
be e.g. superclusters.

\subsection{Without equation of state}
Let us assume that no equation of state is specified and consider the inequality
\begin{align}
      \label{b}
      y(r)\geq \zeta(r) \lp\frac{12\pi P(r)+4\pi\bar{\rho}(r)-\Lambda}
      {12\pi P_{c}+4\pi\rho_{c}-\Lambda}\rp\,.
\end{align}
The functions 
$y(r)$ and $\zeta(P(r))$ are the Buchdahl variables~\cite{Buchdahl:1959}, 
already mentioned above:
\begin{align}
      &\zeta(P(r))=\exp\lp -\int_{P_{c}}^{P(r)}\frac{\dif P}{P+\rho(P)}\rp\,, \nonumber \\
      &y(r)=\sqrt{1-\frac{8\pi}{3}\bar{\rho}(r)r^{2}-\frac{\Lambda}{3}r^{2}}\,.
      \label{b1}
\end{align}
For a constant density distribution one finds 
\begin{align}
      \label{b2}
      \zeta(r)=\frac{P_{c}+\rho}{P(r)+\rho}\,,
\end{align}
and equation~(\ref{b}) becomes an equality 
\begin{align}
      \label{b3}
      \sqrt{1-\frac{8\pi}{3}\bar{\rho}r^{2}-\frac{\Lambda}{3}r^{2}}=
      \frac{P_{c}+\rho}{P(r)+\rho}\lp \frac{12\pi P(r)+4\pi\bar{\rho}(r)-\Lambda}
      {12\pi P_{c}+4\pi\rho_{c}-\Lambda}\rp\,.
\end{align}
Solving for the pressure and using $\rho_{c}=\bar{\rho}=\rho$ we find
\begin{align}
      \label{b4}
      P(r)=\rho \left[\frac{(1-\beta)\lp P_{c}+\rho[1-y(r)]\rp-3P_{c}y(r)}
      {3P_{c}(y(r)-1)+\rho(y(r)[1-\beta]-3)}\right]\,,
\end{align}
where $\beta = \Lambda/4 \pi\rho$ and moreover 
\begin{align}
      \label{b4a}
      P_{c}=\rho(1-\beta) \frac{y(R)-1}{3y(R)-(1-\beta)}\,.
\end{align}
Finiteness of the central pressure implies that the denominator must
be well defined, hence greater than zero. In the simplified constant
density case this implies
\begin{align}
      y(R) > \frac{1}{3}(1-\beta) \,,
      \label{eq:buch.add1}
\end{align}
which is just the Buchdahl inequality~(\ref{eq:buch1}).

To justify why in the constant density case the inequality~(\ref{b}) 
becomes an equality, one can compare with the solution of the TOV-$\Lambda$ 
equation for constant density~\cite{Boehmer:2003uz}. One finds
\begin{align}
      \label{b4a1}
      P(r)=\rho \frac{(\beta-1)+C y(r)}{3-Cy(r)}\,,
\end{align}
where $C$ is a constant of integration which can be fixed with the 
condition $P(R)=0$. This yields
\begin{align}
      \label{b4a2}
      C=\frac{1-\beta}{y(R)}\,.
\end{align}
On the other hand, at the center $r=0$ we have
\begin{align}
      \label{b4a3}
      P_{c}=\rho\frac{(\beta-1)(y(R)-1)}{3y(R)-(1-\beta)}\,,
\end{align}
and therefore the constant of integration becomes
\begin{align}
      \label{b4a4}
      C=\frac{3P_{c}+(1-\beta)\rho}{P_{c}+\rho}\,.
\end{align}
Solving for $y(r)$ from~(\ref{b4a1}) yields 
\begin{align}
      \label{b4a5}
      y(r)=\frac{1}{C}\frac{\rho(1-\beta)+3P(r)}{P(r)+\rho}\,.
\end{align}
If we replace the constant $C$ from~(\ref{b4a4}) we obtain equation~(\ref{b3}) 
and hence~(\ref{b4}). This justifies the conversion from an inequality to an equality 
in~(\ref{b}) for constant density. Thus it is trivially fulfilled 
for constant density solutions. 

\section{General solutions with equation of state}
In view of the above results, we can assume a variable density in order 
to find a more restrictive condition on $\Lambda$.
First of all, inequality~(\ref{b}) is valid for any density profile; 
on the other hand, since it must be fulfilled at any $r$ in order that 
the pressure be a decreasing function of the radius, we can evaluate 
it at the boundary $r=R$, defined by the condition $P(R)=0$, which leads to
\begin{align}
      \label{b5}
      y(R)\geq \zeta(P(R))\lp\frac{4\pi\bar{\rho}_b-\Lambda}
      {12\pi P_{c}+4\pi\rho_{c}-\Lambda}\rp \geq 
      \lp\frac{4\pi\bar{\rho}_b-\Lambda}{12\pi P_{c}+4\pi\rho_{c}-\Lambda}\rp \,.
\end{align}
In this case we can safely assume an equation of state in the form of~(\ref{eos1}), 
so that the central pressure is only a function of the central density. 
The function $\zeta(P)$ can be integrated and becomes a function of the 
central density when evaluated at the boundary
\begin{align}
      \label{b6}
      \zeta(P(R))=\lp\kappa^{\gamma} \rho_{c}^{\gamma(\gamma-1)}+1  \rp^{\frac{\gamma}{\gamma-1}}
      =\zeta(\rho_{c}) \,.
\end{align}
Equation~(\ref{b5}) then becomes a cubic equation for $\Lambda$, namely,
\begin{align}
      \label{b7}
      f(\Lambda)=a\Lambda ^{3}+b\Lambda^{2} +c\Lambda +d \geq 0 \,,
\end{align}
where the coefficients are given explicitly as
\begin{align}
      a = &-\frac{1}{3}R^{3}\,, \qquad
      b = 1+8\pi P_{c}R^{2}-\frac{8\pi}{3}R^{2}\lp\bar{\rho}-\rho_{c}\rp-\zeta^{2}\,, 
      \nonumber \\ \nonumber
      c = &-24\pi P_{c}-48\pi^{2}P_{c}^{2}R^{2} + 64 \pi^{2}\bar{\rho}P_{c}R^{2}
      +8\pi \zeta^{2} \bar{\rho}-8\pi\rho_{c}-32\pi^{2}P_{c}\rho_{c}R^{2} \\ \nonumber
      &+\frac{64}{3}\pi^{2}R^{2}\bar{\rho}\rho_{c}-\frac{16}{3}\pi^{2}R^{2}\rho_{c}^{2}\,, 
      \\ \nonumber
      d = &\,144\pi^{2}P_{c}^{2}-384\pi^{3}P_{c}^{2}\bar{\rho}R^{2}-16\pi^{2}\zeta^{2}\bar{\rho}^{2} 
      + 96\pi^{2}P_{c}\rho_{c}- 256\pi^{3}P_{c}\bar{\rho}\rho_{c}R^{2} \\
      &+16\pi^{2}\rho_{c}^{2}- \frac{128}{3}\pi^{3}\bar{\rho}\rho_{c}^{2}R^{2}\,,
      \label{b8}
\end{align}
and where $\bar{\rho}=\bar{\rho}_b=\bar{\rho}(R)$ is the object's mean density. Thus 
we have the following parameters $R$, $\rho_{c}$, $P_{c}$,
$\bar{\rho}_b$, which can be related as follows: the 
central pressure is connected with the central density through 
the equation of state so that $P_{c}=\kappa \rho_{c}^{\gamma}$. 
The radius of the configuration is related with the mean density at the surface as
\begin{align}
      \label{b9}
      R^{3}=\frac{3M}{4\pi \bar{\rho}_b}\,,
\end{align}
while the mass~(\ref{eq:mass}) is just the volume integral of the density
\begin{align}
      \label{b10}
      M=\int_{0}^{R} 4\pi s^{2}\rho(s)\dif s \,.
\end{align}
We need the density profile and the total mean density 
(together with an equation of state) to solve the cubic 
equation~(\ref{b7}) for $\Lambda$.

It is interesting to further exploit the inequality~(\ref{b5}).
Although it is quite involved to extract information on the 
cosmological constant one can solve it for the
central energy density. This yields 
\begin{align}
      \rho_c \geq \rho_{c,{\rm min}} =
      \frac{\bar{\rho}-\frac{\Lambda}{4\pi}}{y(R)} 
      -3P_c +\frac{\Lambda}{4\pi}\,,
      \label{eq:cen1}
\end{align}
giving a lower bound on the central energy density.
This is not surprising because we have upper bounds
on the boundary mean density. Since the energy density
for astrophysical models is a decreasing function of the radius,
one must find some lower bound, that in particular extends
the boundary mean density.

A clever choice of new variables could be of help in order
to get information on the cosmological constant from~(\ref{b5}).
For example one could use the effective quantities mentioned
in the derivation of equation~(\ref{eq:gen2}). However, the numerator
of~(\ref{eq:cen1}) suggests a definition of type 
$\Delta = \bar{\rho}-\Lambda /4\pi$. In  future we plan 
to explore the astrophysical significance of inequality~(\ref{b5}) 
and other versions thereof.

\section{Inequalities from virial theorem}

In what follows, the non-relativistic virial theorem
is recalled and an upper bound on the cosmological constant
is derived with surprising similarities to those derived
before.

Note that there exist several relativistic formulations of
the virial theorem in literature
\cite{Bonazzola:1973,Bonazzola:1994,Georgiou:1980,Georgiou:2003,Vilain:1979},
however, they are not complete generalizations since some
restrictive assumptions are always needed in their formulation.

The standard non-relativistic virial theorem reads
\begin{align} 
      \label{virial}
      \frac{\dif^2 I_{jk}}{\dif t^2} = 4K_{jk}+2W_{jk} + \ldots \,,
\end{align}
where $I_{jk}$ is the inertial tensor, $K_{jk}$ the kinetic and 
$W_{jk}$ the gravitational potential tensor. The dots in~(\ref{virial}) 
indicate other possible terms due to magnetic field, pressure etc. 
If an external force is exerted on the object, we have to add to the 
right hand side of equation~(\ref{virial}) the term \cite[p.280]{Binney:1987}
\begin{align}   
      \label{external}
      V_{jk}=-\frac{1}{2}\int \rho \left(
      x_k \frac{\partial \Phi_{{\rm ext}}}{\partial x_j}+ 
      x_j \frac{\partial \Phi_{{\rm ext}}}{\partial x_k}\right) \dif^3 x\,,
\end{align}
where $\Phi_{{\rm ext}}$ is the external potential and the case of a 
cosmological constant corresponds to
\begin{align}
      \label{lambdaforce}
      \Phi_{{\rm ext}}=-\frac{1}{6} \Lambda r^2\,.
\end{align}
Therefore the new virial theorem which accounts for the cosmological
constant~\cite{Jackson} takes the form
\begin{align} 
      \label{virial2}
      \frac{\dif^2 I_{jk}}{\dif t^2} = 4K_{jk} + 2W_{jk} 
      +\frac{2}{3} I_{jk}\Lambda + \ldots \,.
\end{align}
This is, in principle, a differential equation for $I_{jk}$ provided
$K_{jk}$ and $W_{jk}$ are given or corresponding differential equations 
in these variables are derived. It is very often more convenient to 
consider a less demanding task by simply noting that the trace $W$ 
of $W_{jk}$ is negative whereas the trace $K$ of $K_{jk}$ is positive
definite. Then the gravitational equilibrium i.e.~$\dif^{2}I_{jk}/\dif t^2=0$ 
yields the inequality
\begin{align} 
      \label{ineqvirial}
      -\frac{1}{3}\Lambda I +\vert W \vert \geq 0\,,
\end{align}
where $I$ denotes the trace of the inertial tensor $I_{jk}$. To appreciate the 
meaning of this inequality we specialize to the case of constant density
$\bar{\rho}=\rho$. It is then easy to show that~(\ref{ineqvirial}) 
takes the form
\begin{align} 
      \label{ineqvirial2}
      8\pi \rho \geq A\Lambda \,,
\end{align}
where the quantity $A$ depends only on the geometry of the object under
consideration. More specifically it reads
\begin{align} 
      \label{A}
      A=\frac{16\pi}{3}\frac{\int r^2 \dif^3x}{\int 
      \frac{\Phi_{{\rm N}}}{\rho} \dif^3x}\,,
\end{align}
where $\Phi_{{\rm N}}$ is the Newtonian part of the non-relativistic gravitational
potential. For spherically symmetric objects one easily calculates $A=2$ and
therefore the virial inequality is simply
\begin{align} 
      \label{ineqvirial3}
      4\pi\rho \geq \Lambda \,.
\end{align}
We already mentioned the similarity between Buchdahl's inequalities
(which are a consequence of hydrostatic equilibrium) and the inequalities
describing the validity of the Newtonian approximation. A second kind of such a
coincidence (if one can call it coincidence) occurs also here with regard
to equation~(\ref{ineqvirial3}), (\ref{eq:newton4}) and~(\ref{eq:gen3}). 
To add to this coincidence we mention that the authors of~\cite{Sussman:2003km} 
by demanding stability of circular orbits derived further the inequality
\begin{align} 
      \label{sussman}
      4\pi\rho \geq 4\pi\rho_c \ge\Lambda \,.
\end{align}
No doubt, all three inequalities originating from different premises have something
in common. The similarity between the three results is not trivial. Certainly,
the hydrostatic and gravitational equilibrium are intuitively related. However,
the hydrostatic equilibrium necessarily involves pressure whereas
(\ref{ineqvirial3}) is derived without using this concept. Indeed, the virial
equation for an object in gravitational equilibrium  with pressure reads
\begin{align} 
      \label{pressure}
      \frac{1}{3}\Lambda I + W +2K +3PV + \ldots=0\,,
\end{align}
where the pressure $P$ is constant over the volume $V$ (note that $P$ 
can have different signs depending whether the pressure is internal 
or external) and the dots indicate the presence of terms related to 
magnetic fields, rotational effects, etc. We think that this issue 
deserves a closer inspection.

With the help of the virial theorem we can show a drastic astrophysical effect
of the cosmological constant. Inequalities like (\ref{ineqvirial}) and
(\ref{ineqvirial2}) are useful to check if a given astrophysical system is in 
gravitational equilibrium without the knowledge of the kinetic tensor or,
which is the same, the average velocity $\langle v^2 \rangle$ 
of the components of the system. It is often of utmost interest to know
this velocity. The knowledge of the shape of the object, its
density profile and $\Lambda$ allows one to calculate
\begin{align} 
      \label{meanv}
      \langle v^2 \rangle =\frac{\vert W \vert}{M} 
      -\frac{8\pi}{3}\frac{\rho_{{\rm vac}}}{M}I\,.
\end{align}
To appreciate the effect of $\Lambda$ let us assume a constant density and the shape of the astrophysical object to be an ellipsoid. 
Then the $W$ and $I$
can be calculated analytically~\cite{Binney:1987}. For the inertial integral $I$ we get
\begin{align} 
      \label{I}
      I=\rho \frac{4}{15}a_1a_2a_3\left(a_1^2 +a_2^2 +a_3^2 \right)\,,
\end{align}
where $a_i$ are the axes of the ellipsoid.  The mean velocity can be 
now written as
\begin{align} 
      \label{meanv2}
      \langle v^2 \rangle_{{\rm ellipsoid}} =\frac{32 \pi}{45}\rho\rho_{{\rm vac}}
      a_1a_2a_3\left(a_1^2+a_2^2+a_3^2\right)\left (\frac{3\pi}{4}
      \Gamma_{{\rm ellipsoid}} -1\right)\,.
\end{align}
For the oblate case ($a_1=a_2 > a_3$, $e=\sqrt{1-a_3^2/a_1^2}\,$)
one calculates
\begin{align} 
      \label{oblate}
      \Gamma_{{\rm oblate}}=\frac{\left (\frac{a_3}{a_1} \right)}
      {1 +\frac{1}{2}\left (\frac{a_3}{a_1} \right)^2}\frac{\sin^{-1}e}{e}\,.
\end{align}
The prolate case ($a_1=a_2< a_3$, $e=\sqrt{1-a_1^2/a_3^2}$)
gives
\begin{align} 
      \label{prolate}
      \Gamma_{{\rm prolate}}=\frac{\left(\frac{a_1}{a_3}\right)^3}
      {1+2\left(\frac{a_1}{a_3}\right)^2}\frac{\ln\left(
      \frac{1+e}{1-e}\right)}{e}\,. 
\end{align}
The triaxial case ($a_1 > a_2 >a_3$, 
$\theta =\cos^{-1}(a_3/a_1)$, 
$k=\sqrt{(a_1^2 -a_2^2)/(a_1^2 -a_3^2)}$) is simply
\begin{align} 
      \label{triaxial}
      \Gamma_{{\rm triaxial}}=\frac{a_1 a_2 F(\theta, k)}{a_1^2+a_2^2+a_3^2}\,,
\end{align}
where $F(\theta, k)$ is the elliptic integral of the first kind.
Note now that for a flattened prolate ellipsoid we can approximate
\begin{align} 
      \label{prolate2}
      \Gamma_{{\rm prolate}}\simeq\left(\frac{a_1}{a_3}\right)^3 
      \ln\left(\frac{1+e}{1-e}\right)\,.
\end{align}
Since the nowadays preferred value of $\rho_{{\rm vac}}$ is
$(0.7-0.8)\rho_{{\rm crit}}$ we can say that if the 
constant $\rho/\rho_{{\rm crit}}$ is, say, $10^{3}$, it 
suffices for the ellipsoid to have the ratio $a_1/a_3 \sim 10$ in order 
that the mean velocity of its components approaches zero. 
This is valid always under the assumption that the object 
is in gravitational equilibrium. 
This effect is due to the relatively large cosmological constant. In general
we can say that in flattened astrophysical systems in gravitational
equilibrium, the mean velocity gets affected by the cosmological constant.
The denser the system, the bigger should be the deviation from spherical
symmetry to have a sizable effect.
It is interesting if such an effect can be observed in reality which would 
confirm the existence of $\Lambda$. One can paraphrase this also by looking at the
results from a different
perspective. If we are certain that a given astrophysical object is in
gravitational equilibrium, then equation (52) and (53) would put a stringent bound
on $\Lambda$ in the case of strong deviation from spherical symmetry. 

From the last equations we can infer that the power of equilibria concepts with
$\Lambda$ lies in analyzing relatively low density astrophysical objects whose
shapes deviate from the spherical symmetry. It is evident that this can easily be
done using the virial theorem. We have shown that for spherically symmetric objects
the virial theorem and the hydrostatic equilibrium yield surprisingly similar results.
It would be therefore desirable to have a tool to probe the hydrostatic equilibrium 
with $\Lambda$ for arbitrarily shaped bodies as well. The full examination of this aspect
of hydrostatic equilibrium goes beyond the scope of the present paper.
We can, however, outline a direction in which such an investigation
might proceed following~\cite{Lombardi:2001}.
The basic equations for the Newtonian concept of hydrostatic equilibrium are
\begin{align}
      \label{Equil}
      \nabla P = -\rho\nabla\Phi\,, \qquad
      \nabla^2 \Phi = 4\pi \rho -\Lambda\,,
\end{align}
together with~(\ref{eos1}) as the equation of state. Defining $u=\rho/\rho_0$,
$\mbox{\boldmath $\xi$}=s{\bf x}$ with $s^2=4\pi\rho^{2-\gamma}/\kappa \gamma$ 
we obtain for $u=u(\mbox{\boldmath $\xi$})$
\begin{align}
      \label{emden}
      (1-\gamma)\left (\frac{\nabla_{\xi}u}{u}\right)^2 + 
      \nabla_{\xi}^2 \ln u = 
      a_{\Lambda} u^{1-\gamma} -u^{2-\gamma}\,,
\end{align}
which is a generalized Lane-Emden equation. For $\Lambda=0$ we 
have to set $a_{\Lambda}=0$ whereas the case $\Lambda \neq 0$ 
requires $a_{\Lambda}=2$. Furthermore for $\Lambda =0$, $\rho_0$ is arbitrary,
but is fixed to be $\rho_0=\rho_{\rm vac}$ for non-zero cosmological constant. 
We will examine the consequences of equation~(\ref{emden}) elsewhere. 
However, here we can already mention a difference between the cases 
$\Lambda=0$ and $\Lambda \neq 0$ due to the fact that the cosmological 
constant sets a scale for the density. Consider the case $\gamma=1$. 
Then with $\gamma=1$ and for $\Lambda=0$ we have a scaling property in 
the following sense: if $u(\mbox{\boldmath $\xi$})$ is a solution of the  
equation, so is $\lambda^2 u(\lambda \mbox{\boldmath $\xi$})$~\cite{Chandrasekhar:1967}. 
This similarity property is lost if $a_{\Lambda}=2$. 

Coming back to the virial theorem we note that
interestingly one can eliminate the cosmological constant if we address the
question whether an astrophysical object is in gravitational equilibrium.
Let the system which is in gravitational equilibrium be described by
$I_1$, $W_1$ and $K_1$, etc. This system serves us as a reference system. 
Defining $\epsilon_{21}\equiv I_{2}/I_{1}=1/\epsilon_{12}$ and provided
$\Lambda \neq 0$, a second system (denoted by the subscript 2) is
also in gravitational equilibrium if the following relation is fulfilled
\begin{align} 
      \label{comparison}
      \epsilon_{21}=\frac{W_2+ 2K_2+\Xi_{2}}{W_1 + 2 K_1+\Xi_{1}}\,,
\end{align}
where $\Xi_{i}$ contains terms related to pressure, magnetic field, etc. 
On the other hand, if we know two systems bounded by an external medium at a
constant pressure $P_{{\rm ext}}$ (this means that $\Xi_{i}=-3P_{{\rm ext}}V_i$ 
assuming that there is no magnetic field) are in equilibrium, 
we can solve for the pressure and obtain
\begin{align} 
      \label{expressure}
      P_{{\rm ext}}=\frac{1}{3V_1}
      \left(\frac{V_2}{V_1}-\epsilon_{21} \right)^{-1}
      \left[W_1\left(\frac{W_2}{W_1}-\epsilon_{21}\right)
      +2K_1\left(\frac{K_2}{K_1} -\epsilon_{21}\right)\right]\,.
\end{align}
Note that the salient and necessary assumption for equations~(\ref{comparison}) 
and~(\ref{expressure}) is a non-zero cosmological constant $\Lambda$. 
With these equations we can put in relation two systems in equilibrium 
or solve for the variable common in both systems.
Solving for the kinetic energy of system $2$ we have
\begin{align}
      \label{ex7}
      K_{2}=\frac{1}{2}\lp 2 \epsilon_{21} K_{1}+
      \epsilon_{21} W_{1}-W_{2} +\epsilon_{21} \Xi_{1}-\Xi_{2}\rp\,.
\end{align}
Since this is a positive definite quantity, we find an inequality for $W_{2}$
\begin{align}
      \label{ex8}
      |W_{2}|\geq \epsilon_{21} \lp \vert W_{1}\vert 
      -2K_{1}-\Xi_{1}+\frac{1}{\epsilon_{21}}\Xi_{2}\rp\,,
\end{align}
which gives an estimate of the left-hand side. 

As an example, we assume again that the system $1$ is in 
steady state and that both systems are in equilibrium with an 
external pressure $P_{{\rm ext}}$, as before. 
Under these conditions we can solve for the pressure and find
\begin{align}
      \label{ex9}
      P_{{\rm ext}}\leq \frac{\vert W_{1}\vert}{3V_{1}} 
      \lp\frac{\vert W_{2}\vert}{\vert W_{1}\vert}-\epsilon_{21} \rp 
      \lp\epsilon_{21}-\frac{V_{2}}{V_{1}} \rp^{-1}\,,
\end{align}
which can serve as a lower bound for the external pressure.
The question that rises here is, what can we identify as the reference system? 
On the one hand, one could argue that at every different astrophysical scale 
there must be a standard system, so that one must find, for instance, 
a standard galaxy in order to obtain information about a certain other galaxy. 
The exploration of these interesting aspects will be the subject of
further research. We conclude this section by the remark that the virial 
theorem with non-zero $\Lambda$ has also been successfully 
applied in cosmology~\cite{Barrow:1989}.

\section{Scales of the cosmological constant}

To appreciate the orders of magnitude estimate we reinstall 
here the Newtonian coupling constant $G$ which we put to $1$ 
in the preceding sections. The cosmological constant sets 
scales for density, length, mass and time. The density
scale in equation (1) indicates the so-called coincidence 
problem. The latter can be formulated as the question as to 
why we are living in an epoch in which $\rho_{\rm vac}
\sim \rho_{crit}$. More precisely, since $\rho_{\rm crit}$ 
is epoch dependent, the coincidence $\rho_{\rm vac} \sim \rho_{crit}$ 
is a time coincidence of a sharp transition from small 
ratio $\Omega_{\Lambda}$ to $\Omega_{\Lambda} \sim 1$~\cite{Frampton:2004nh}.
We can reformulate this coincidence in terms of length, 
noting that
\begin{align}
      \label{scaleL}
      r_{\Lambda} =\frac{1}{\sqrt{\Lambda}}=0.476\,H_0^{-1}=
      1.43 \times 10^3 h_0^{-1}{\rm Mpc}\,,
\end{align}
which is of cosmological order of magnitude and close to the 
radius of the visible universe. The combination
\begin{align}
      \label{scaleM}
      M_{\Lambda}=\frac{1}{G\sqrt{\Lambda}}=3.61 
      \times 10^{22} h_0^{-1}M_{\odot}
      \sqrt{\frac{\rho_{\rm crit}}{\rho_{\rm vac}}}\,,
\end{align}
is close to the mass of the universe. As noted in~\cite{Nowakowski:2000dr}       
this is not trivial since $\Lambda$ as a constant could be either smaller or bigger 
than what is assumed today. We could equally well live in an epoch
where the universe is smaller or bigger or the mass of the universe 
is not close to $M_{\Lambda}$. Interestingly, both such scales play 
an important role in the considerations of the validity of the weak 
field approximation and in the hydrostatic equilibrium in the form of 
Buchdahl inequalities (see the discussion in section~\ref{se:buch}). Hereby 
the differences of the numerical factors are of order one. Had we lived 
in a different universe or a different epoch of our universe, the 
above scales would appear only in the Newtonian limit and the Buchdahl 
inequalities, which by itself is a remarkable technical coincidence.
However, in our universe these scales appear in all considerations: 
cosmology, hydrostatic equilibrium and Newtonian limit. Notice also 
that one of the equations of motion of a test body in the Schwarzschild 
metric with $\Lambda$ is the relation between the affine parameter 
$\lambda$ and the time $t$ which reads
\begin{align}
      \label{horizonts}
      \left(1-2\frac{r_S}{r}-\frac{1}{3}\frac{r^2}{r^2_{\Lambda}}\right)
      \frac{dt}{d\lambda}=E={\rm const.}\,,
\end{align}
with $r_S=GM$. The case $E=0$ sets the limit of validity of the 
coordinate system. This means that the expression on the left hand 
side of~(\ref{horizonts}) is zero which in turn leads to the solutions 
$r_{\rm min}=2r_S$ and $r_{\rm max}=\sqrt{3}r_{\Lambda}$. Albeit the 
cases $r=r_{\rm min}\, , r_{\rm max}$ are singular points due to the 
choice of coordinates, the condition above tells us also that we should 
not go beyond the horizon of a black hole and also not beyond the 
horizon of the universe. The latter emerges only because of the above 
coincidence of length mentioned above. In the same context, i.e.,
considering the motion in a Schwarzschild spacetime the scale $R$ defined by
\begin{align}
      \label{R}
      R^3= 3r_Sr_{\Lambda}^2\,,
\end{align}
not only yields astrophysically relevant length scales (this is due to 
the combination of a small and large quantity),
but also has the meaning of the largest possible radius within 
which bound orbits in the Schwarzschild metric
are possible~\cite{Balaguera-Antolinez:2005}.

Curiously, not all scales are of cosmological or astrophysical relevance 
even though they have the right order of magnitude. The time scale
connected to $\Lambda$ is
\begin{align} 
      \label{time}
      T_{\Lambda}= 4.65 h_0^{-1}\,{\rm Gyr}\,,
\end{align}
which is close to the age of our solar system, but should be considered 
as an accidental coincidence.

The scale $m_{\Lambda}=\sqrt{\Lambda}=3.1 \times 10^{-42}\, {\rm GeV}$
has no connection to particle physics, but interestingly establishes 
one more of Dirac's large numbers in the form 
$m_p/(G\sqrt{\Lambda}) \sim 3.2 \times 10^{41}$ where $m_p$ is the 
nucleon's mass.

\section{Summary}
In this work we investigated the astrophysical limits on the cosmological 
constant for spherically symmetric bodies. These limits emerged 
from examinations of hydrostatic and gravitational equilibrium as 
well as from constraints on the validity of the Newtonian limit. 
Although the resulting inequalities are based on 
quite different premises, they do resemble each other to an extent
which let us conjecture a deeper underlying principle or connection.
As a side result we conjecture the existence of a new class
of static and spherically symmetric perfect fluid solutions 
with the cosmological constant. By coincidence the length and mass
scales set by $\Lambda$ appear as the horizon and mass of the universe and 
simultaneously as the limits of validity of the weak field expansion and the
hydrostatic equilibrium. This adds to the puzzle of the coincidence problem.

In this context we note that the inequalities based on the hydrostatic
equilibrium were derived in the general relativistic framework whereas 
the virial inequalities required only a Newtonian approximation. We suspect 
that general relativistic virial equations might be
a more suitable tool to compare the hydrostatic equilibrium with the 
virial one.

Already the Newtonian virial theorem is a powerful equation 
in many situations where the astrophysical object is not spherically symmetric.
Indeed, flattened, large and diluted objects cannot be in gravitational
equilibrium if the cosmological constant is as large as claimed 
nowadays. It is therefore of some interest to extend the consideration of the
hydrostatic equilibrium allowing for an arbitrary shape of the objects.
Unfortunately, the general relativistic treatment of arbitrarily shaped 
objects is quite out of reach at the moment, but we have indicated how to do that
in the Newtonian approach. This will be covered elsewhere.
Similarly, some of our results, like the inequality (43) still await its
practical exploitation.

The inequality $2\rho \ge \rho_{\rm vac}$ which we found from various
considerations is valid provided that the object is
spherically shaped and in equilibrium. Assuming this and taking into account
that $\rho$ can be at the most of the order of the background density i.e.
$\rho_{\rm crit}$, this is not a weak inequality. Indeed, it comes very close to the
preferred $\rho_{\rm vac} \sim 0.7 \rho_{\rm crit}$.

In this work we have laid down the basics of equilibria concepts 
including a cosmological constant. As indicated in the paper one 
should continue this study by examining hydrostatic equilibrium 
for non-spherical objects and generalize the equilibria concepts 
also to other theories of the Dark Energy~\cite{Mota:2004pa}. 

We can look upon many inequalities derived in the main text by 
imposing the condition of equilibrium from two different angles.
Assuming a known value of $\Lambda$ the inequalities limit the 
possible values of the density of the astrophysical object if the 
latter is assumed in equilibrium. If we know
that an object is in equilibrium and assuming we can infer
the mean density of such an object from some other considerations, 
then the same inequalities put a limit on the cosmological constant.

\section*{Acknowledgements}
One of the authors CGB wishes to thank Herbert Balasin, Daniel Grumiller
and Wolfgang Kummer for their constant support. This work has partly 
been supported by project P-15449-N08 of the Austrian Science Foundation
(FWF) and is part of the research project Nr. 01/04 
{\it Quantum Gravity, Cosmology and Categorification} 
of the Austrian Academy of Sciences (\"OAW) and the National
Academy of Sciences of Ukraine (NASU).

\addcontentsline{toc}{section}{References}
\bibliographystyle{unsrt}
\bibliography{/home/boehmer/spinors/bib/review,/home/boehmer/spinors/bib/add_ref}

\begin{thebibliography}{10}

\bibitem{Krauss:1998}
L.~M. Krauss.
\newblock The {E}nd of the {A}ge {P}roblem, and the {C}ase for a {C}osmological
  {C}onstant {R}evisited.
\newblock {\em Astrophys. J.}, 501:461, 1998.

\bibitem{Perlmutter:1999jt}
S.~Perlmutter, M.~S. Turner, and M.~J. White.
\newblock Constraining dark energy with sne ia and large-scale structure.
\newblock {\em Phys. Rev. Lett.}, 83:670--673, 1999.

\bibitem{Ratra:1987rm}
B.~Ratra and P.~J.~E. Peebles.
\newblock Cosmological consequences of a rolling homogeneous scalar field.
\newblock {\em Phys. Rev.}, D37:3406, 1988.

\bibitem{Wetterich:1987fm}
C.~Wetterich.
\newblock Cosmology and the fate of dilatation symmetry.
\newblock {\em Nucl. Phys.}, B302:668, 1988.

\bibitem{Kamenshchik:2001cp}
A.~Yu. Kamenshchik, U.~Moschella, and V.~Pasquier.
\newblock An alternative to quintessence.
\newblock {\em Phys. Lett.}, B511:265--268, 2001.

\bibitem{Kremer:2002hz}
G.~M. Kremer and F.~P. Devecchi.
\newblock Viscous cosmological models and accelerated universes.
\newblock {\em Phys. Rev.}, D67:047301, 2003.

\bibitem{Kremer:2003ev}
G.~M. Kremer.
\newblock Irreversible processes in a universe modelled as a mixture of a
  chaplygin gas and radiation.
\newblock {\em Gen. Rel. Grav.}, 35:1459--1466, 2003.

\bibitem{Kremer:2003vs}
G.~M. Kremer.
\newblock Cosmological models described by a mixture of van der waals fluid and
  dark energy.
\newblock {\em Phys. Rev.}, D68:123507, 2003.

\bibitem{Kremer:2004tw}
G.~M. Kremer and D.~S.~M. Alves.
\newblock Acceleration field of a universe modeled as a mixture of scalar and
  matter fields.
\newblock {\em Gen. Rel. Grav.}, 36:2039--2051, 2004.

\bibitem{Nojiri:2003}
S.~Nojiri and S.~D. Odintsov.
\newblock Modified gravity with negative and positive powers of curvature:
  {U}nification of inflation and cosmic acceleration.
\newblock {\em Phys. Rev.}, D68:123512, 2003.

\bibitem{Zeldovich:1967gd}
Y.~B. Zel'dovich.
\newblock Cosmological constant and elementary particles.
\newblock {\em JETP Lett.}, 6:316, 1967.

\bibitem{Carneiro:2002iv}
S.~Carneiro.
\newblock Holography and the cosmic coincidence.
\newblock 2002.
\newblock {\it gr-qc/0206064}.

\bibitem{Tegmark:2000qy}
M.~Tegmark, M.~Zaldarriaga, and A.~J.~S. Hamilton.
\newblock Towards a refined cosmic concordance model: joint 11- parameter
  constraints from cmb and large-scale structure.
\newblock {\em Phys. Rev.}, D63:043007, 2001.

\bibitem{Matthews:1988}
R.~A.~J. Matthews.
\newblock Dirac's coincidences 60 years on.
\newblock {\em Astronomy and Geophysics}, 39:19--20, 1988.

\bibitem{Eddington:1931}
A.~Eddington.
\newblock {On the Value of the Cosmical Constant}.
\newblock {\em Royal Society of London Proceedings Series A}, 133:605--615,
  October 1931.

\bibitem{MenaMarugan:2001qn}
Mena M., Guillermo A., and S.~Carneiro.
\newblock Holography and the large number hypothesis.
\newblock {\em Phys. Rev.}, D65:087303, 2002.

\bibitem{Krauss:2003yb}
L.~M. Krauss.
\newblock The state of the universe: Cosmological parameters 2002.
\newblock In {\em {E}{S}{O}-{C}{E}{R}{N} {S}ymposium on {A}stronomy,
  {C}osmology and {F}undamental {P}hysics}, March 2003.

\bibitem{Perlmutter:1998zf}
S.~Perlmutter et~al.
\newblock Discovery of a supernova explosion at half the age of the universe
  and its cosmological implications.
\newblock {\em Nature}, 391:51--54, 1998.

\bibitem{Riess:2004nr}
A.~G. Riess et~al.
\newblock Type ia supernova discoveries at {$z>1$} from the hubble space
  telescope: Evidence for past deceleration and constraints on dark energy
  evolution.
\newblock {\em Astrophys. J.}, 607:665--687, 2004.

\bibitem{Allen:2004cd}
S.~W. Allen, R.~W. Schmidt, H.~Ebeling, A.~C. Fabian, and L.~van Speybroeck.
\newblock Constraints on dark energy from chandra observations of the largest
  relaxed galaxy clusters.
\newblock 2004.
\newblock {\it astro-ph/0405340}.

\bibitem{Kunz:2004ry}
M.~Kunz and B.~A. Bassett.
\newblock A tale of two distances.
\newblock 2004.

\bibitem{Gamow:1939}
G.~Gamov and E.~Teller.
\newblock {O}n the {O}rigin of {G}reat {N}ebulae.
\newblock {\em Phys. Rev.}, 55:654--657, 1939.

\bibitem{North:1990}
J.~D. North.
\newblock {\em The {M}easure of the {U}niverse. {A} study of {M}odern
  {C}osmology}.
\newblock Dover Publications, 1990.

\bibitem{Maldacena:1998re}
J.~M. Maldacena.
\newblock The large {$N$} limit of superconformal field theories and
  supergravity.
\newblock {\em Adv. Theor. Math. Phys.}, 2:231--252, 1998.

\bibitem{Volkov:1996qj}
M.~S. Volkov, N.~Straumann, George~V. Lavrelashvili, M.~Heusler, and
  O.~Brodbeck.
\newblock Cosmological analogues of the {B}artnik--{M}c{K}innon solutions.
\newblock {\em Phys. Rev.}, D54:7243--7251, 1996.

\bibitem{Brihaye:2004qr}
Y.~Brihaye and B.~Hartmann.
\newblock Spherically symmetric solutions of a (4+n)-dimensional
  {E}instein-{Y}ang-{M}ills model with cosmological constant.
\newblock 2004.

\bibitem{Boehmer:2002gg}
C.~G. B{\"o}hmer.
\newblock General relativistic static fluid solutions with cosmological
  constant.
\newblock 2002.
\newblock {\it gr-qc/0308057, unpublished Diploma Thesis}.

\bibitem{Boehmer:2003uz}
C.~G. B{\"o}hmer.
\newblock Eleven spherically symmetric constant density solutions with
  cosmological constant.
\newblock {\em Gen. Rel. Grav.}, 36:1039--1054, 2004.

\bibitem{Stuchlik:2000}
Z.~Stuchl{\'{\i}}k.
\newblock Spherically symmetric static configurations of uniform density in
  spacetimes with a non-zero cosmological constant.
\newblock {\em Acta Physica Slovaca}, 50:219--228, 2000.

\bibitem{Weyl:1919}
H.~Weyl.
\newblock {\"U}ber die statischen kugelsymmetrischen {L}{\"o}sungen von
  {E}insteins \flqq kosmologischen\frqq\ {G}raviatationsgleichungen.
\newblock {\em Physikalische Zeitschrift}, 20:31--34, 1919.

\bibitem{Nowakowski:2000dr}
M.~Nowakowski.
\newblock {The consistent Newtonian limit of Einstein's gravity with a
  cosmological constant}.
\newblock {\em Int. J. Mod. Phys.}, D10:649--662, 2001.

\bibitem{Sussman:2003km}
R.~A. Sussman and X.~Hernandez.
\newblock On the {N}ewtonian limit and cut--off scales of isothermal dark
  matter halos with cosmological constant.
\newblock {\em Mon. Not. Roy. Astron. Soc.}, 345:871, 2003.

\bibitem{Nowakowski:2001zw}
M.~Nowakowski, J.-C. Sanabria, and A.~Garcia.
\newblock {Gravitational equilibrium in the presence of a positive cosmological
  constant}.
\newblock {\em Phys. Rev.}, D66:023003, 2002.

\bibitem{Boehmer:2003iv}
C.~G. B{\"o}hmer.
\newblock The {E}instein static universe with torsion and the sign problem of
  the cosmological constant.
\newblock {\em Class. Quant. Grav.}, 21:1119--1124, 2004.

\bibitem{Dimakis:1985jb}
A.~Dimakis and F.~Mueller-Hoissen.
\newblock Solutions of the {E}instein-{C}artan-{D}irac equations with vanishing
  energy momentum tensor.
\newblock {\em J. Math. Phys.}, 26:1040, 1985.

\bibitem{VanNieuwenhuizen:1981ae}
P.~van Nieuwenhuizen.
\newblock Supergravity.
\newblock {\em Phys. Rept.}, 68:189--398, 1981.

\bibitem{Weinberg:1972}
S.~Weinberg.
\newblock {\em {G}ravitation and cosmology}.
\newblock John Wiley \& Sons, 1972.

\bibitem{Oppenheimer:1939ne}
J.~R. Oppenheimer and G.~M. Volkoff.
\newblock On massive neutron cores.
\newblock {\em Phys. Rev.}, 55:374--381, 1939.

\bibitem{Tolman:1939jz}
R.~C. Tolman.
\newblock Static solutions of {E}instein's field equations for spheres of
  fluid.
\newblock {\em Phys. Rev.}, 55:364--373, 1939.

\bibitem{Ehlers:1991}
J.~Ehlers.
\newblock {\em {T}he {N}ewtonian limit of general relativity}, page 112.
\newblock In Ferrarese \cite{Ferrarese:1991}, 1991.

\bibitem{Ferrarese:1991}
G.~Ferrarese, editor.
\newblock {\em {C}lassical {M}echanis and {R}elativity: {R}elationship and
  {C}onsistency}.
\newblock American Institute of Physics, 1991.

\bibitem{Buchdahl:1959}
H.~A. Buchdahl.
\newblock General relativistic fluid spheres.
\newblock {\em Phys. Rev.}, 116:1027--1034, 1959.

\bibitem{Rendall:1991hg}
A.~D. Rendall and B.~G. Schmidt.
\newblock Existence and properties of spherically symmetric static fluid bodies
  with a given equation of state.
\newblock {\em Class. Quant. Grav.}, 8:985--1000, 1991.

\bibitem{Stephani:1967}
H.~Stephani.
\newblock {\"U}ber {L}{\"o}sungen der {E}insteinschen {F}eldgleichungene, die
  sich in einen f{\"u}nfdimensionalen flachen {R}aum einbetten lassen.
\newblock {\em Commun. Math. Phys.}, 4:137--142, 1967.

\bibitem{Beig:2000yf}
R.~Beig and B.~G. Schmidt.
\newblock Time-independent gravitational fields.
\newblock {\em Lect. Notes Phys.}, 540:325--372, 2000.

\bibitem{Mak:2001gg}
M.~K. Mak, Jr. Dobson, P.~N., and T.~Harko.
\newblock Maximum mass-radius ratio for compact general relativistic objects in
  {S}chwarzschild-de {S}itter geometry.
\newblock {\em Mod. Phys. Lett.}, A15:2153--2158, 2000.

\bibitem{Bonazzola:1973}
S.~{Bonazzola}.
\newblock {The Virial Theorem in General Relativity}.
\newblock {\em Astrophy. J.}, 182:335--342, 1973.

\bibitem{Bonazzola:1994}
S.~Bonazzola and E.~Gourgoulhon.
\newblock A virial identity applied to relativistic stellar models.
\newblock {\em Class. Quant. Grav.}, 11:1775--1784, 1994.

\bibitem{Georgiou:1980}
A.~Georgiou.
\newblock A virial theorem for general relativistic charged fluids.
\newblock {\em J. Phys.}, A13:3751--3759, 1980.

\bibitem{Georgiou:2003}
A.~Georgiou.
\newblock A virial theorem for rotating charged perfect fluids in general
  relativity.
\newblock {\em Class. Quant. Grav.}, 20:359--367, 2003.

\bibitem{Vilain:1979}
C.~Vilain.
\newblock {Virial theorem in general relativity - Consequences for stability of
  spherical symmetry}.
\newblock {\em Astrophy. J.}, 227:307--318, 1979.

\bibitem{Binney:1987}
J.~Binney and S.~Tremaine.
\newblock {\em {G}alactic {D}ynamics}.
\newblock Princeton University Press, 1987.
\newblock (Princeton Series in Astrophysics).

\bibitem{Jackson}
J.~C.~Jackson.
\newblock The dynamics of clusters of galaxies in universes with non-zero cosmological constant, and the virial theorem mass discrepancy.
\newblock {\em Mon. Not. Roy. Astron. Soc.} 148:249--260, 1970.

\bibitem{Lombardi:2001}
M.~Lombardi and G.~Bertin.
\newblock Boyle's law and gravitational instability.
\newblock {\em {Astronomy \& Astrophysics}}, 375:1091, 2001.

\bibitem{Chandrasekhar:1967}
S.~Chandrasekhar.
\newblock {\em An Introduction to the Study of Stellar Structure}.
\newblock Dover, 1967.

\bibitem{Barrow:1989}
J.~D. Barrow and G.~Gotz.
\newblock Newtonian no-hair theorems.
\newblock {\em Class. Quant. Grav.}, 6:1253--1265, 1989.

\bibitem{Frampton:2004nh}
P.~H. Frampton.
\newblock Dark energy: A pedagogic review.
\newblock 2004.

\bibitem{Balaguera-Antolinez:2005}
A.~Balaguera-Antol{\'{\i}}nez, C.~G. {B\"ohmer}, and M.~Nowakowski.
\newblock Sales set by the cosmological constant.
\newblock 2005.
\newblock In preperation.

\bibitem{Mota:2004pa}
D.~F. Mota and C.~van~de Bruck.
\newblock On the spherical collapse model in dark energy cosmologies.
\newblock {\em {Astronomy \& Astrophysics}}, 421:71, 2004.

\end{thebibliography}

\end{document}